\documentclass[aps,prx,reprint,superscriptaddress,twocolumn,showpacs,amsmath,amssymb,floatfix]{revtex4-2}

\usepackage[utf8]{inputenc}
\usepackage[T1]{fontenc}
\usepackage[english]{babel}
\usepackage{natbib}

\usepackage{graphicx}
\usepackage{dcolumn}
\usepackage{bm}
\usepackage{gensymb}
\usepackage{multirow}

\usepackage{epstopdf}
\usepackage{float}
\usepackage{subfigure}
\usepackage{tabularx}
\usepackage{color}
\usepackage{siunitx}
\usepackage{xr}
\usepackage{amsmath} 
\usepackage{mathtools}
\usepackage{braket}
\usepackage[normalem]{ulem}
\usepackage{natbib}
\usepackage[caption=false]{subfig}
\usepackage{afterpage}
\usepackage{enumerate}

\makeatletter
\newcommand*{\rom}[1]{\expandafter\@slowromancap\romannumeral #1@}
\makeatother

\begin{document}

\newcommand{\lsy}[1]{{\color{red} #1}}

\newcommand{\sr}[1]{{\color{magenta} #1}}
\newcommand{\beq}{\begin{equation}}
\newcommand{\eeq}{\end{equation}}
\newcommand{\vS}{\vec{S}}
\newcommand{\ud}{\mathrm{d}}

\title{Altermagnetism and Strain Induced Altermagnetic Transition in Cairo Pentagonal Monolayer}
\author{Shuyi Li}
\affiliation{Department of Physics, University of Florida, Gainesville, FL 32611, USA}

\author{Yu Zhang}
\affiliation{Department of Physics, University of Florida, Gainesville, FL 32611, USA}

\author{Adrian Bahri}
\affiliation{Department of Physics, University of Florida, Gainesville, FL 32611, USA}

\author{Xiaoliang Zhang}
\affiliation{Department of Physics, University of Florida, Gainesville, FL 32611, USA}

\author{Chunjing Jia}
\email[Correspondence e-mail address: ]{chunjing@phys.ufl.edu}
\affiliation{Department of Physics, University of Florida, Gainesville, FL 32611, USA}

\date{\today}

\begin{abstract}
Altermagnetism, a recently discovered class of magnetic order characterized by vanishing net magnetization and spin-splitting band structures, has garnered significant research attention. In this work, we introduce a novel two-dimensional system that exhibits $g$-wave altermagnetism and undergoes a strain-induced transition from $g$-wave to $d$-wave altermagnetism. This system can be realized in an unconventional monolayer Cairo pentagonal lattice, for which we present a realistic tight-binding model that incorporates both magnetic and non-magnetic sites. Furthermore, we demonstrate that non-trivial band topology can emerge in this system by breaking the symmetry that protects the spin-polarized nodal points. Finally, \emph{ab initio} calculations on several candidate materials, such as FeS$_2$ and Nb$_2$FeB$_2$, which exhibit symmetry consistent with the proposed tight-binding Hamiltonian, are also presented. These findings open new avenues for exploring spintronic devices based on altermagnetic systems.
\end{abstract}

\maketitle

\vspace{8mm}
\noindent
\MakeUppercase{\textbf{Introduction}}

Recently, altermagnetism has attracted growing research interest due to its unconventional behavior, which is distinct from traditional collinear ferromagnetism and antiferromagnetism~\cite{AMth1,AMth2,AMth3,AMth4,AMth5,AMth6,AMsplit1,AMsplit2}. In altermagnet, magnetic moments form an antiferromagnetic-like order with zero net magnetization, while it exhibits energy splitting between states with opposite spins, similar to ferromagnets. A variety of materials have been proposed or confirmed to exhibit altermagnetism through first-principles calculations and experiments, including RuO$_2$~\cite{RuO21,RuO22,RuO23}, MnTe~\cite{MnTe1,MnTe2,MnTe3,MnTe4}, FeSb$_2$~\cite{FeSb21}, and CrSb~\cite{CrSb1}. The unique electronic structure in altermagnets leads to numerous interesting effects and potential applications, such as spin-splitting torque phenomena~\cite{spintorque1,spintorque2}, unconventional superconductivity~\cite{AMSC1,AMSC2,AMSC3,AMSC4,AMSC5,AMSC6,AMSC7,AMSC8}, and distinct variants of Hall effect~\cite{AMhall1,AMhall2,AMhall3,AMhall4,AMhall5,AMhall6,AMhall7,AMweyl1,AMweyl2}. In addition, altermagnets hold great promise for spintronic applications due to their large spin-splitting and robustness against magnetic field perturbations~\cite{AMth2,spintronics1,spintronics2}. 

One of the most fundamental questions in the study of altermagnetic materials is how to understand the origin of this emergent behavior, which is essential for predicting new physical properties and exploring potential applications. At the microscopic level, several mechanisms for the emergence of altermagnetism have been proposed, including the interplay between magnetic and nonmagnetic atoms and the anisotropic ordering of local orbitals~\cite{AMSC3,Morder}. Thus, identifying realistic tight-binding models from a microscopic perspective is crucial for fully understanding this unconventional phenomenon. While several studies have explored altermagnetism using effective models, only a few have investigated the microscopic origin of altermagnetic properties~\cite{AMSC3,Morder,AMmodel1,AMmodel2,AMmodel3}. 

In our work, we investigate the microscopic theory for the origin of altermagnetism on a novel two-dimensional pentagonal structure~\cite{penta1}, known as the Cairo Pentagon. We uncover the mystery of altermagnetism on the Cairo pentagonal lattice at a microscopic level by constructing a simple but realistic tight-binding model containing both magnetic and non-magnetic atoms, and demonstrate that the interplay between them plays a key role in the origin of altermagnetism. Another remarkable feature of the pentagonal lattice is its sensitivity to strain, which induces strong in-plane anisotropy~\cite{penta2,penta3,penta4}. This anisotropy alters the spin-lattice symmetry, thereby impacting the structure of spin-splitting. Calculated band structure using the tight-binding model shows that this strong dependence on strain will lead to a transition between $g$-wave and $d$-wave altermagnetism. In addition, we find that this pentagonal altermagnet hosts symmetry protected polarized nodal points, and we give examples demonstrating how breaking this symmetry could gap out these nodal points and lead to non-trivial topological bands. Finally, we examine two candidate materials FeS$_2$ and Nb$_2$FeB$_2$ through \emph{ab initio} calculations, the results successfully reproduce the altermagnetism and its transition under strain as we expect. 

Our results represent the first demonstration of strain tuning to achieve different altermagnetic orders in realistic systems, along with a microscopic understanding of the underlying mechanisms.  This paves a new avenue for the design and application of strain-tuned spintronic devices based on altermagnetism.\\

\noindent
\MakeUppercase{\textbf{Results}}
\\
\textbf{Cairo pentagonal lattice structure and model Hamiltonian}

\begin{figure}[htbp]
\centering
\includegraphics[width=1.0\linewidth]{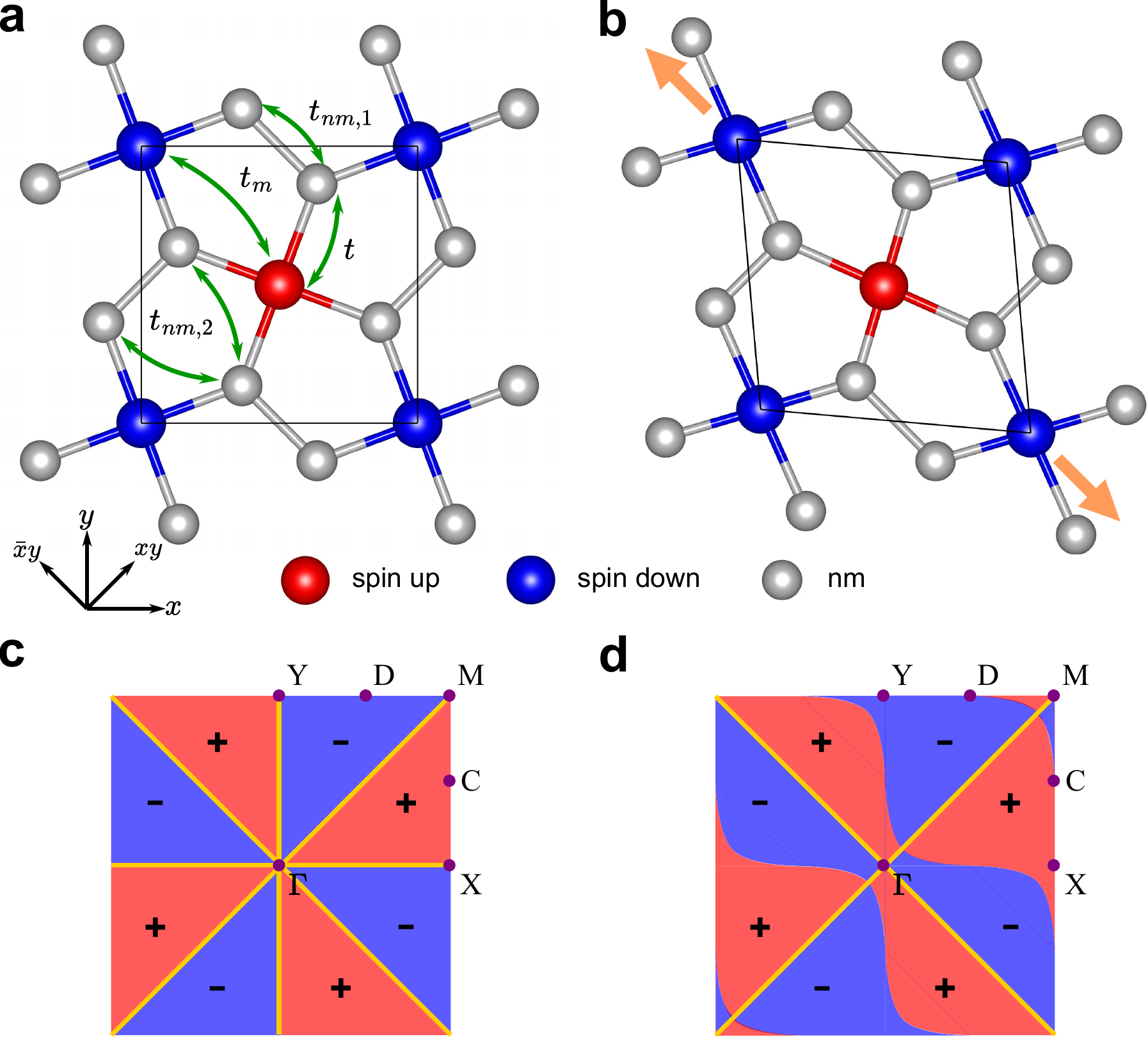}
\caption{\textbf{Cairo pentagonal lattice structure and its electronic energy bands.} \textbf{a} A schematics of the pentagonal lattice with space group $P4/mbm$ and the tight-binding model parameters. The magnetic sites with spin-up, spin-down, and non-magnetic (nm) sites are represented by red, blue, and gray colors. $t$, $t_m$, $t_{nm,1}$ and $t_{nm,2}$ represent the hopping strengths between the nearest-neighbor magnetic and non-magnetic sites, the nearest-neighbor magnetic sites, the first nearest-neighbor non-magnetic sites, and the second nearest-neighbor non-magnetic sites, respectively. \textbf{b} A schematics of the same pentagonal lattice under diagonal strain along $xy$ or $\bar{x}y$. \textbf{c} and \textbf{d} Visualization of the higher energy bands from the pair of spin-split bands on this lattice, without and with strain, in the first Brillouin zone. The bands with spin up and down are represented by red and blue, respectively. The yellow lines represent the spin-degenerate nodal lines that cross the $\Gamma$ point. A number of high-symmetry momentum points are highlighted.}
\label{Fig:model}
\end{figure}

We consider a Cairo pentagonal lattice with space group $P4/mbm$, shown in Fig.~\ref{Fig:model}a. Due to the presence of non-magnetic sites, the opposite spin sublattices can not be mapped to each other by the combination of time-reversal with translation or inversion, which makes it an ideal platform to investigate the relationship between atomic interplay and altermagnetism. 

The lattice is centrosymmetric due to the four-fold rotation symmetry $C_{4z}$. In one primitive unit cell, there are two magnetic atoms with opposite collinear spins and four non-magnetic atoms. The presence of non-magnetic sites breaks time-reversal symmetry while preserving symmetries $\{C_{2\perp}||\mathcal{M}_{x}\}$, $\{C_{2\perp}||\mathcal{M}_{y}\}$, $\{C_{2\perp}||\mathcal{M}_{xy}\}$ and $\{C_{2\perp}||\mathcal{M}_{\bar{x}y}\}$ relating two magnetic sublattices, where $C_{2\perp}$ is the 180\degree~rotation operator around an axis perpendicular to the spins, $\mathcal{M}_{x}$, $\mathcal{M}_{y}$, $\mathcal{M}_{xy}$ and $\mathcal{M}_{\bar{x}y}$ are the mirror operators about the $x$ axis, $y$ axis, and the diagonals $xy$ and $\bar{x}y$. Due to symmetry considerations, the pair of electronic states with opposite spins split at general $k$ points but remain degenerate along $k_y=0$, $k_x=0$, $k_x=k_y$ and $k_x=-k_y$ in the first Brillouin zone, as shown in Fig.~\ref{Fig:model}c. This unconventional spin-splitting ensures that the magnetic pentagonal crystal is classified as an altermagnet.\\

To model the electronic structure in this pentagonal lattice, we consider a tight-binding Hamiltonian including both magnetic and non-magnetic sites in Eq.~(\ref{Eq-C4model}):
\begin{equation}\label{Eq-C4model}
\begin{aligned}
    H&=-\sum_{\langle i,j\rangle,\sigma}t_{ij}c_{i\sigma}^{\dagger}c_{j\sigma}-J\sum_{i\in\text{m},\sigma,\sigma'}\mathbf{S}_i\cdot c_{i\sigma}^{\dagger}\bm{\sigma}_{\sigma\sigma'}c_{i\sigma'}\\
    &+(\epsilon_{m}-\mu)\sum_{i\in m,\sigma}c_{i\sigma}^{\dagger}c_{i\sigma}+(\epsilon_{nm}-\mu)\sum_{i\in nm,\sigma}c_{i\sigma}^{\dagger}c_{i\sigma},
\end{aligned}
\end{equation}
where $c_{i\sigma}^{(\dagger)}$ is the annihilation (creation) operator of an electron at site $i$ with spin $\sigma=\uparrow,\downarrow$, and the hopping strength between electrons at sites $i$ and $j$ is described by $t_{ij}$. In our model, we take into account the hopping $t$ between the nearest-neighbor magnetic site and non-magnetic site, the hopping $t_m$ between the nearest-neighbor magnetic sites, and the hopping $t_{nm,1}$ ($t_{nm,2}$) between the first (second) nearest-neighbor non-magnetic sites. $J$ is the coupling between the electronic spins and localized magnetic moments $\mathbf{S}_i$, and $\bm{\sigma}$ is the Pauli matrix. In this work, we set $\mathbf{S}_i=(0,0,Se^{i\mathbf{q}\mathbf{r}_i})$ with magnetic wave vector $\mathbf{q}=(\pi/a,\pi/a)$ and $S=1$. The on-site energy of magnetic and non-magnetic sites and their chemical potential are denoted by $\epsilon_{m}$, $\epsilon_{nm}$ and $\mu$.\\

We obtain the electronic band structure by diagonalizing the Hamiltonian matrix at each momentum point (see details in the supplementary material). Without spin-orbit coupling, there are no interactions between electrons with opposite spins, and thus spin $\sigma$ is a good quantum number. Finally, the Hamiltonian in Eq.~(\ref{Eq-C4model}) becomes
\begin{equation}
H=\sum_{n,\mathbf{k},\sigma}E_{n,\sigma}(\mathbf{k})f_{n,\mathbf{k},\sigma}^{\dagger}f_{n,\mathbf{k},\sigma},
\end{equation}
where $f_{n,\mathbf{k},\sigma}~(f_{n,\mathbf{k},\sigma}^{\dagger})$ is a fermionic annihilation (creation) operator and $E_{n,\sigma}(\mathbf{k})$ represents the corresponding energy dispersion of the $n$th band with spin $\sigma$ ($n=1,\dots,6$, with energy increasing from low to high). For simplicity, we set $t$ as the energy unit in the following discussion.\\

\noindent
\textbf{Altermagnetic band splittings of lattice with $C_{4z}$ symmetry}
\\

\begin{figure}[htbp]
\centering
\includegraphics[width=1.0\linewidth]{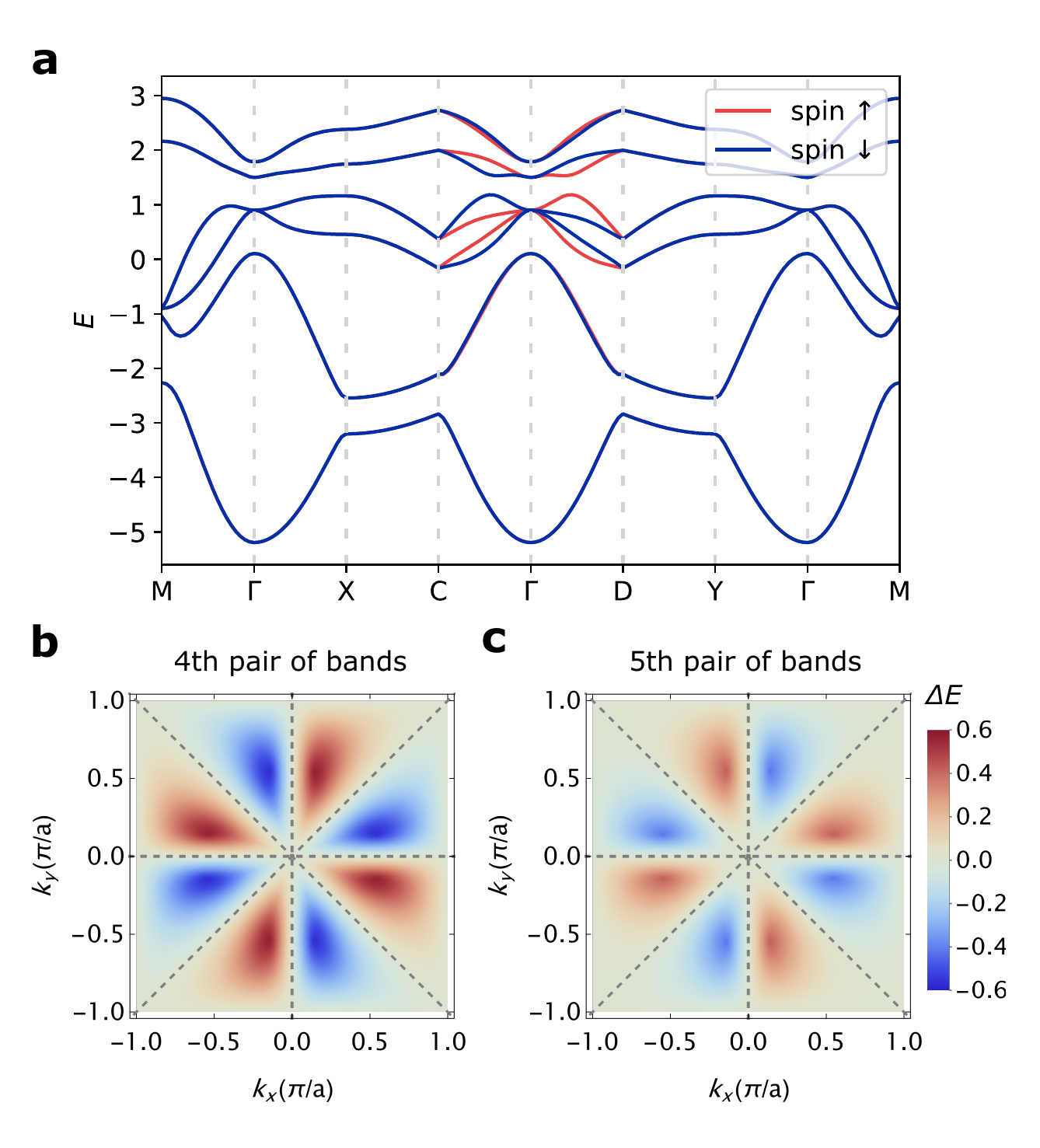}
\caption{\textbf{Band dispersion of pentagonal lattice Hamiltonian with $C_{4z}$ symmetry.} \textbf{a} The electronic band structure along the chosen $\mathbf{k}$ path in the first Brillouin zone. The parameters used are $t_m=0.2$, $t_{nm,1}=0.9$, $t_{nm,2}=0.6$, $J=1$, $\epsilon_m=\epsilon_{nm}=0$ and $\mu=0$. The spin up and spin down bands are shown in red and blue, respectively. \textbf{b} and \textbf{c} False color plots of the spin-splitting energy $\Delta E_n(\mathbf{k})=E_{n,\uparrow}(\mathbf{k})-E_{n,\downarrow}(\mathbf{k})$ for the 4th and 5th pair of bands in the first Brillouin zone. Four dashed lines represent the spin-degenerate nodal lines that cross the $\Gamma$ point.}
\label{Fig:C4}
\end{figure}

The band dispersion of the above pentagonal lattice Hamiltonian exhibits $g$-wave altermagnetism, as shown in Fig.~\ref{Fig:C4}. Fig.~\ref{Fig:C4}a shows six pairs of bands with inverse spin-splitting along $\Gamma$-$C$ and $\Gamma$-$D$. Fig.~\ref{Fig:C4}b and \ref{Fig:C4}c plot the energy spin-splitting $\Delta E_n(\mathbf{k})=E_{n,\uparrow}(\mathbf{k})-E_{n,\downarrow}(\mathbf{k})$ for the 4th and 5th pair of bands. In the first Brillouin zone, there are four spin-degenerate nodal lines, $k_y=0$, $k_x=0$, $k_x=k_y$ and $k_x=-k_y$, crossing the $\Gamma$ point, as expected from the symmetry. Additionally, the spin-splitting $\Delta E_n(\mathbf{k})$ for isolated pairs of bands near the $\Gamma$ point is directly proportional to $k_xk_y(k_x^2 - k_y^2)$ (see details in the supplementary material). All these results illustrate that the tight-binding model successfully realizes $g$-wave altermagnetism.

\begin{figure}[htbp]
\centering
\includegraphics[width=1.0\linewidth]{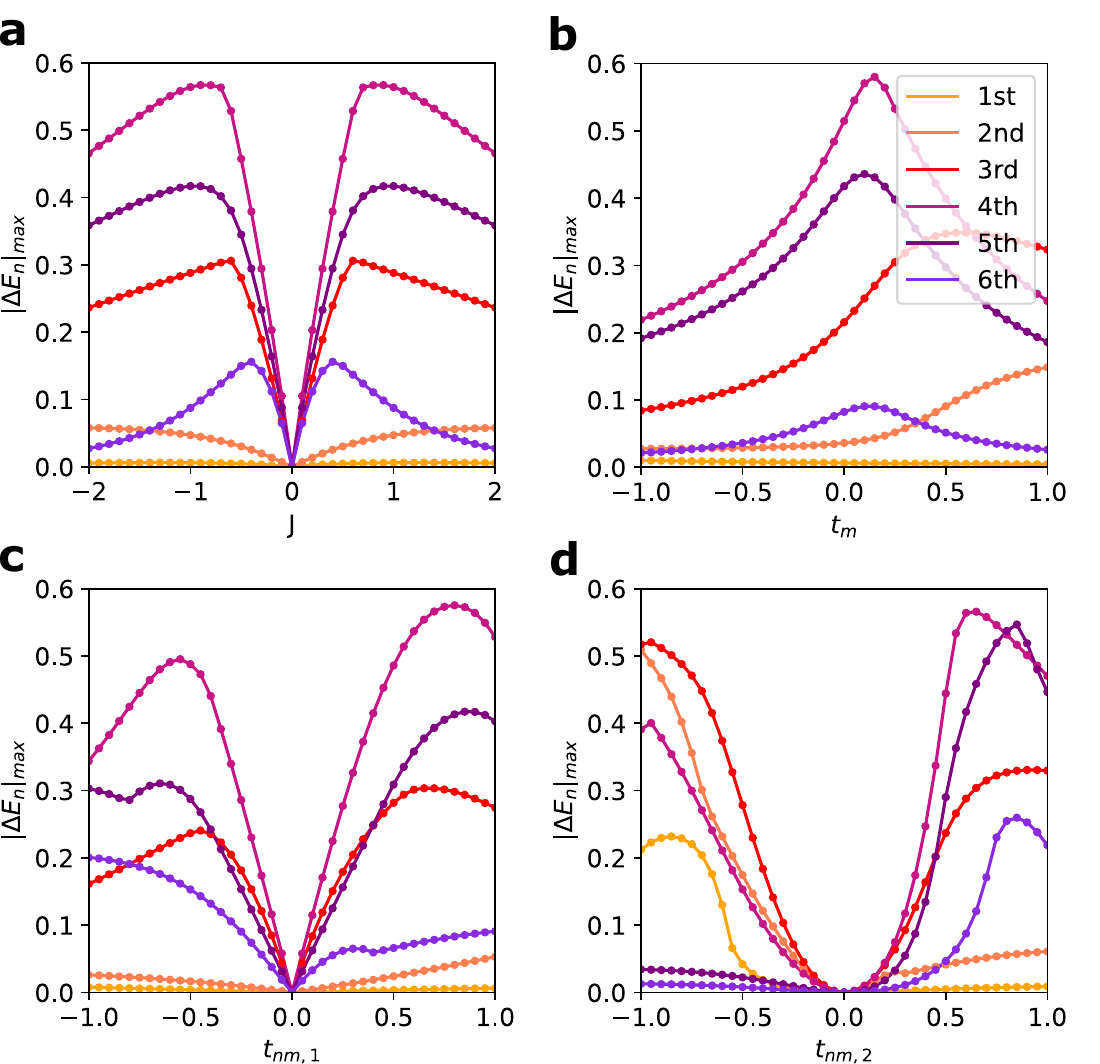}
\caption{\textbf{Dependence of spin-splitting energy on tight-binding model parameters.} Plots of the maximal energy splitting for the $n$-th pair of bands, $|\Delta E_n|_{max}$, as a function of \textbf{a} $J$, \textbf{b} $t_m$, \textbf{c} $t_{nm,1}$, and \textbf{d} $t_{nm,2}$, while other parameters are fixed as $t_m=0.2$, $t_{nm,1}=0.9$, $t_{nm,2}=0.6$, $J=1$, $\epsilon_m=\epsilon_{nm}=0$ and $\mu=0$. Different colors represent the band pairs, ordered from lower to higher energy.}
\label{Fig:split}
\end{figure}

To understand the origin of $g$-wave altermagnetism in this pentagonal lattice, we investigate the dependence of spin-splitting on parameters of the tight-binding Hamiltonian in Eq.~(\ref{Eq-C4model}). The maximal energy splitting of the $n$-th pair of bands $|\Delta E_n|_{max}=Max[|\Delta E_n(\mathbf{k})|]$ as a function of $J$, $t_m$, $t_{nm,1}$ and $t_{nm,2}$ is summarized in Fig.~\ref{Fig:split}. The spin-splitting exhibits a complex dependence on the parameters of the tight-binding model, and the different pairs of electronic bands show significant distinctions. It is evident that the hopping $t$ between magnetic and non-magnetic sites, and the coupling $J$ between the electronic spins and localized magnetic moments are necessary for the system to exhibit altermagnetism. If $t=0$, the magnetic sites form a normal N\'eel antiferromagnet; while if $J=0$, there is no distinction between the sub-Hamiltonians for spin up and down.

In addition to these factors, all energy splittings $|\Delta E_n|_{max}$ vanish when either $t_{nm,1}$ or $t_{nm,2}$ becomes zero. In these cases, the Hamiltonian in Eq.~(\ref{Eq-C4model}) always exhibits band degeneracy $E_{n,\sigma}(k_x,k_y) = E_{n,\sigma}(k_x,-k_y) = E_{n,\sigma}(-k_x,k_y)$, as well as symmetries $\{C_{2\perp}||\mathcal{M}_x\}$ and $\{C_{2\perp}||\mathcal{M}_y\}$ (see details in the supplementary material), which lead to band degeneracy $E_{n,\uparrow}(\mathbf{k}) = E_{n,\downarrow}(\mathbf{k})$ at all $\mathbf{k}$ points. The above reasoning indicates that the hopping terms between non-magnetic sites are also crucial for the emergence of altermagnetic spin splitting in the pentagonal lattice.

\begin{figure}[htbp]
\centering
\includegraphics[width=1.0\linewidth]{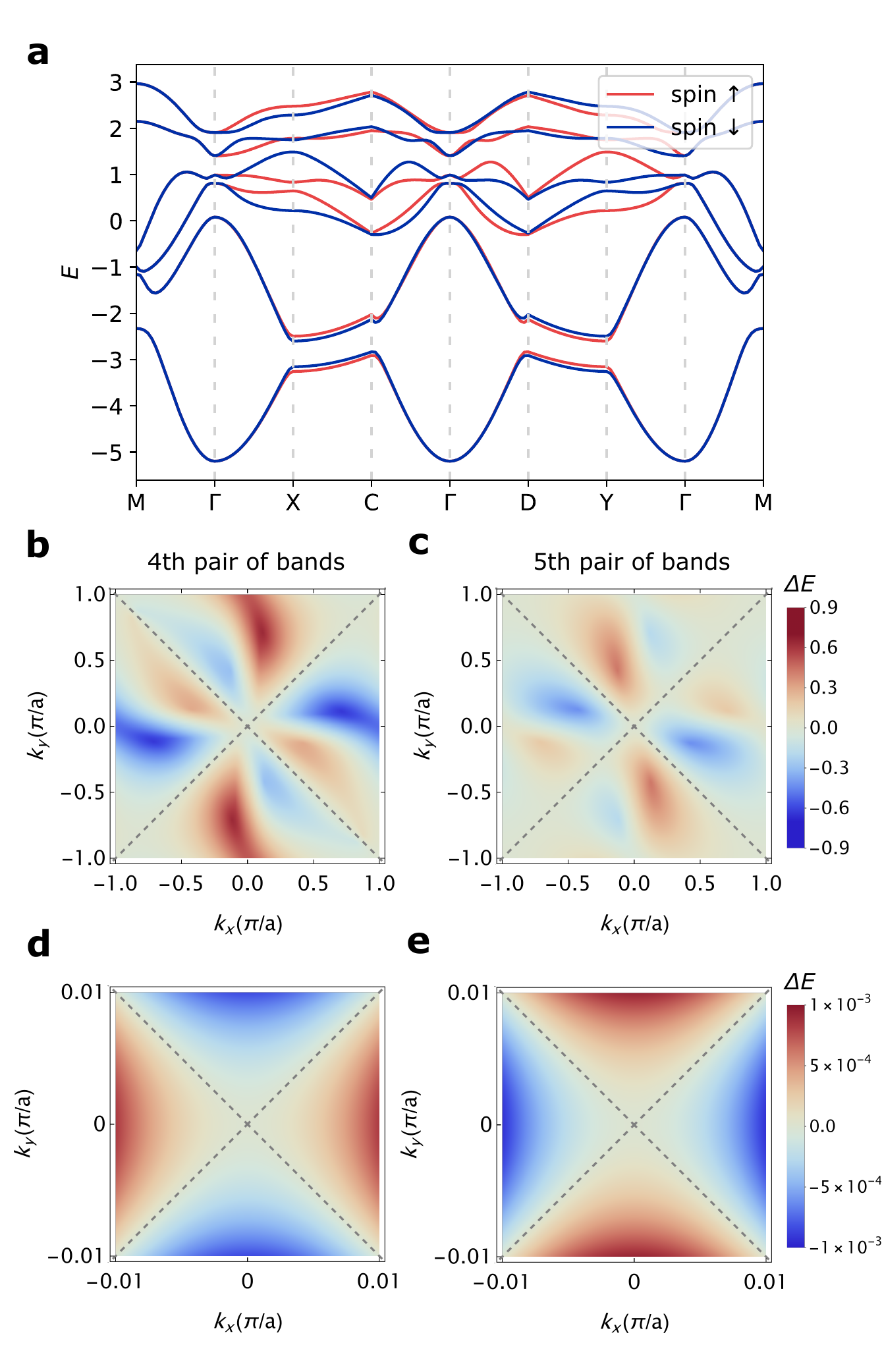}
\caption{\textbf{Band dispersion of pentagonal lattice Hamiltonian under $xy$ and $\bar{x}y$ strains.} \textbf{a} The electronic band structure along the chosen $\mathbf{k}$ path in the first Brillouin zone. The parameters used are $\delta=0.1$, $t_m=0.2$, $t_{nm,1}=0.9$, $t_{nm,2}=0.6$, $J=1$, $\epsilon_m=\epsilon_{nm}=0$ and $\mu=0$. The bands of spin up and down are labeled by red and blue. \textbf{b} and \textbf{c} False color plots of the spin-splitting energy $\Delta E_n(\mathbf{k})=E_{n,\uparrow}(\mathbf{k})-E_{n,\downarrow}(\mathbf{k})$ for the 4th and 5th pair of bands in the first Brillouin zone. \textbf{d} and \textbf{e} Plots of $\Delta E_n(\mathbf{k})$ near the $\Gamma$ points, within $|k_x|\leq 10^{-2}\pi/a$ and $|k_y|\leq 10^{-2}\pi/a$.}
\label{Fig:C2}
\end{figure}

\noindent
\textbf{Altermagnetic band splittings of lattice under $xy$ and $\bar{x}y$ strains}
\\

Next, we study the impact of strain along the diagonal directions $xy$ and $\bar{x}y$ on the altermagnetism of the Cairo pentagonal lattice, as shown in Fig.~\ref{Fig:model}b. After applying the strain, a significant in-plane anisotropy emerges due to the lattice's sensitivity to it. This anisotropy breaks the lattice symmetry $C_{4z}$, and the symmetries $\{C_{2\perp}||\mathcal{M}_{x}\}$ and $\{C_{2\perp}||\mathcal{M}_{y}\}$ relating two magnetic sublattices, while preserving two other symmetries $\{C_{2\perp}||\mathcal{M}_{xy}\}$ and $\{C_{2\perp}||\mathcal{M}_{\bar{x}y}\}$. The change in symmetry results in a transformation in the structure of unconventional spin-splitting, as shown in Fig.~\ref{Fig:model}d.

To incorporate the effect of strain into the tight-binding Hamiltonian, we introduce anisotropic hopping between sites along the $xy$ and $\bar{x}y$ directions. Consequently, the hopping term in Eq.~(\ref{Eq-C4model}) is modified as

\begin{equation}\label{Eq-C2model}
    H_{S}^{hp}=-(1+\delta)\sum_{\langle i,j\rangle_{xy}}t_{ij}c_{i\sigma}^{\dagger}c_{j\sigma}-(1-\delta)\sum_{\langle i,j\rangle_{\bar{x}y}}t_{ij}c_{i\sigma}^{\dagger}c_{j\sigma},
\end{equation}
where the anisotropy $\delta$ reflects the strength of strain.

The band dispersion relations are shown in Fig.~\ref{Fig:C2}a. Unlike the previous case, six pairs of bands now exhibit inverse spin splitting along $X\mbox{-}\Gamma\mbox{-}C$ and $Y\mbox{-}\Gamma\mbox{-}D$. The energy spin-splittings $\Delta E_n(\mathbf{k})$ shown in Fig.~\ref{Fig:C2}b and Fig.~\ref{Fig:C2}c demonstrate that there are only two spin-degenerate nodal lines, $kx = \pm ky$, crossing the $\Gamma$ point, while the two former nodal lines are shifted away. The change in nodal lines is consistent with the symmetry breaking induced by diagonal strain, while the two preserved symmetries $\{C_{2\perp}||\mathcal{M}_{xy}\}$ and $\{C_{2\perp}||\mathcal{M}_{\bar{x}y}\}$ enforce the existence of the remaining two nodal lines. In Fig.~\ref{Fig:C2}d and Fig.~\ref{Fig:C2}e, we zoom in on the above plots near the $\Gamma$ point, where the spin-splitting $\Delta E_n(\mathbf{k})$ is directly proportional to $k_x^2 - k_y^2$ (see details in the supplementary material) and thus exhibits $d$-wave symmetry. Consequently, as a result of the applied strain, the system transitions from a $g$-wave to a $d$-wave altermagnetism in the neighborhood of the $\Gamma$ point.

\begin{figure}[htbp!]
\centering
\includegraphics[width=1.0\linewidth]{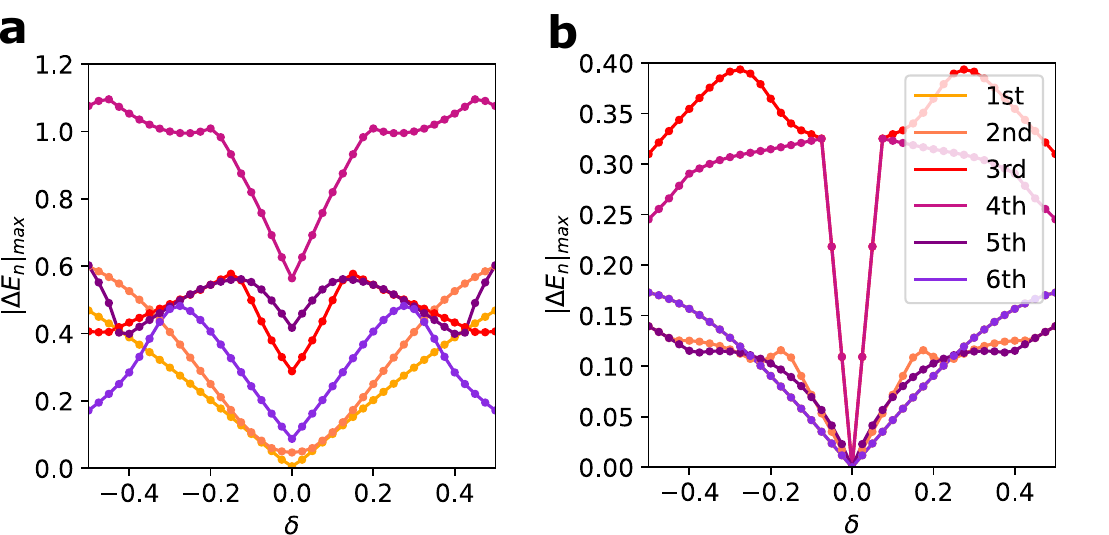}
\caption{\textbf{Dependence of spin-splitting on strain strength.} Plots of the maximal energy splitting for the $n$-th pair of bands, $|\Delta E_n|_{max}$, as a function of anisotropy $\delta$ under \textbf{a} $t_{nm,2}=0.6$ and \textbf{b} $t_{nm,2}=0$, with other parameters being $t_m=0.2$, $t_{nm,1}=0.9$, $t_{nm,2}=0.6$, $J=1$, $\epsilon_m=\epsilon_{nm}=0$ and $\mu=0$. Different colors represent the band pairs, ordered from lower to higher energy.}
\label{Fig:splitdelta}
\end{figure}

Additionally, we examine the dependence of spin-splitting on the strain strength. Fig.~\ref{Fig:splitdelta}a illustrates the maximal spin-splitting $|\Delta E_n|_{max}$ as a function of anisotropy $\delta$. The sensitivity of the band splitting to strain varies among different pairs of bands. Within the range of $|\delta|\leq 0.5$, the maximum splitting can reach slightly higher than the energy level of $t$. For comparison, Fig.~\ref{Fig:splitdelta}b depicts the case where the next-nearest-neighbor hopping between the magnetic and nonmagnetic sites $t_{nm,2}$ is zero. As discussed earlier, the unstrained system exhibits normal spin-degeneracy in this case. Upon applying diagonal strain, the system transitions from a normal antiferromagnetic phase to a $d$-wave altermagnetic phase, with the maximal spin-splitting showing an almost linear dependence on $\delta$ for small values of $\delta$.\\

\noindent
\textbf{Polarized Nodal Points and Band Topology}
\\

\begin{figure*}[htbp]
\centering
\vspace{-4mm}
\includegraphics[width=1.0\linewidth]{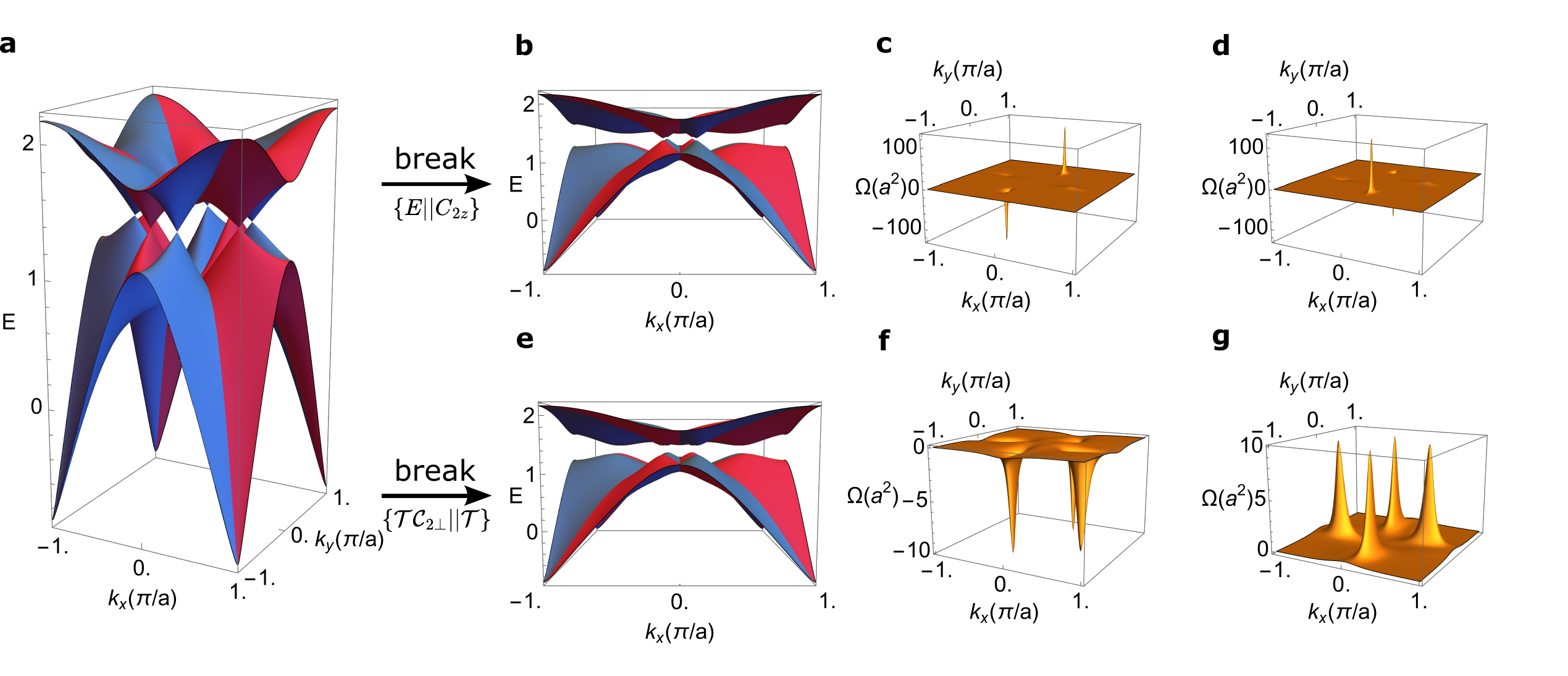}
\caption{\textbf{Band structure and Berry curvature after gapping out the polarized nodal points.} \textbf{a} Eight spin polarized nodal points formed by band crossing between the 4th and 5th pairs of bands. \textbf{b} The electronic band structure for the 4th and 5th pairs of bands under inversion symmetry $\{E||C_{2z}\}$ breaking with $t_1=0.2$, and \textbf{c}, \textbf{d} the corresponding Berry curvature $\Omega(\mathbf{k})$ of the 5th pair of spin-up and spin-down bands. \textbf{e} The electronic band structure for the 4th and 5th pairs of bands under inversion symmetry $\{\mathcal{T}C_{2\perp}||\mathcal{T}\}$ breaking with $t_2=0.2$, and \textbf{f}, \textbf{g} the corresponding Berry curvature $\Omega(\mathbf{k})$ of the 5th pair of spin-up and spin-down bands. The other parameters in the calculations are $t_m=0.2$, $t_{nm,1}=0.9$, $t_{nm,2}=0.6$, $J=1$, $\epsilon_m=\epsilon_{nm}=0$ and $\mu=0$.}
\label{Fig:BC}
\end{figure*}

In addition to the nodal lines formed by spin-degeneracy, several spin-polarized nodal points are present in this altermagnetic pentagonal lattice. As shown in Fig.~\ref{Fig:BC}a, eight spin-polarized nodal points emerge at the band crossings of the 4th and 5th bands with the same spin. These nodal points are protected by the symmetry $\{\mathcal{T}C_{2\perp}||\mathcal{T}C_{2z}\}$, with each nodal point carrying a $\pi$ Berry phase, where $\mathcal{T}$ represents the time-reversal operator. Thus, nontrivial band topology can be achieved by breaking this symmetry. Here, we explore two mechanisms that gap out the spin-polarized nodal points using toy models: breaking the lattice inversion symmetry $\{E||C_{2z}\}$ through the anisotropic hopping term between magnetic atoms, represented by
\begin{equation}\label{Eq-H1}
    H_1=\sum_{i\in (m\uparrow),\sigma}\sum_{\bm{\xi}}(t_1c^{\dagger}_{i,\sigma}c_{i+\bm{\xi}\sigma}-t_1c^{\dagger}_{i\sigma}c_{i-\bm{\xi},\sigma})+h.c.,
\end{equation}
and breaking the lattice time-reversal symmetry $\{\mathcal{T}C_{2\perp}||\mathcal{T}\}$ through the complex hopping term between magnetic atoms, represented by
\begin{equation}\label{Eq-H2}
    H_2=-\sum_{i\in (m\uparrow),\sigma}(-1)^{\sigma}\sum_{\bm{\xi}}i(t_2c^{\dagger}_{i,\sigma}c_{i+\bm{\xi}\sigma}+t_2c^{\dagger}_{i\sigma}c_{i-\bm{\xi},\sigma})+h.c.,
\end{equation}
where $\bm{\xi}=\frac{1}{2}a\hat{x} \pm \frac{1}{2}a\hat{y}$.

The impact of breaking the symmetry $\{E||C_{2z}\}$ using Eq.~(\ref{Eq-H1}) is illustrated in Fig.~\ref{Fig:BC}b. All spin-polarized nodal points are gapped out, while the spin-degeneracy nodal lines are preserved. We plot the Berry curvature $\Omega(\mathbf{k})$ of the 5th pairs of bands with opposite spins after inversion symmetry breaking in Fig.~\ref{Fig:BC}c and Fig.~\ref{Fig:BC}d, where the peaks of $\Omega(\mathbf{k})$ appear at the former nodal points. In this case, the Berry curvature satisfies $\Omega(\mathbf{k})=-\Omega(-\mathbf{k})$ for all $\mathbf{k}$ points, resulting in Chern number $\mathcal{C}_{\sigma}=0$. In contrast, the band structure after $\{\mathcal{T}C_{2\perp}||\mathcal{T}\}$ symmetry breaking using Eq.~(\ref{Eq-H2}) and the Berry curvature of the 5th pairs of bands are shown in Fig.~\ref{Fig:BC}e, Fig.~\ref{Fig:BC}f and Fig.~\ref{Fig:BC}g. In each band, this symmetry-breaking term in Eq.~(\ref{Eq-H2}) leads to an equivalent contribution from four nodal points. This mechanism mirrors the behavior seen in Haldane's model, making the 5th pair of bands topologically non-trivial with a Chern number $\mathcal{C}_{\sigma}=\mp2$.\\

\begin{figure*}[htbp]
\centering
\includegraphics[width=0.8\linewidth]{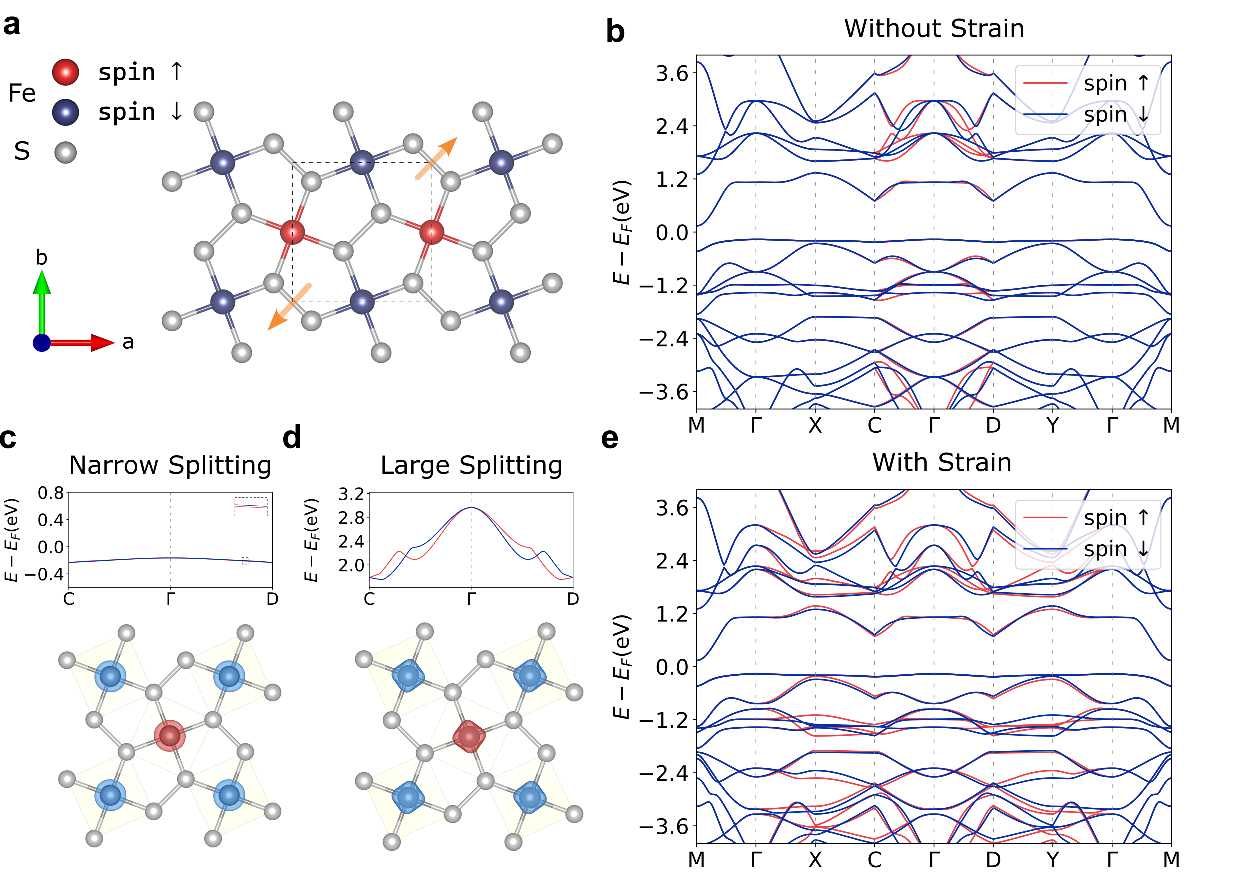}
\caption{\textbf{Crystal structure and electronic bands of FeS$_2$.} \textbf{a} Crystal structure of FeS$_2$. The red and blue colors of Fe atoms indicate the directions of their localized magnetic moments, while the orange arrows represent the direction of diagonal strain. \textbf{b} The electronic band structure of FeS$_2$ without strain. \textbf{c} and \textbf{d} Spin density isosurfaces for two pairs of bands without strain, illustrating narrow and large spin-splitting, respectively (the inset zooms in on energy by
a factor of $5$). \textbf{e} The electronic band structure of FeS$_2$ with diagonal strain, where the angle $\gamma$ between lattice parameters $a$ and $b$ is tuned to $\gamma = 88\degree$.}
\label{Fig:FeS2}
\end{figure*}

\noindent
\textbf{Candidate Materials}
\\

A two-dimensional form of pyrite, FeS$_2$, has been predicted to exhibit a Cairo pentagonal structure with antiferromagnetic order~\cite{pentagonDFT}. It belongs to space group $P4/mbm$ (No.127) and point group $D_{4h}$. Fig.~\ref{Fig:FeS2}a shows the crystal structure, where each Fe atom is surrounded by four S atoms. To examine the magnetic ground state, we perform \emph{ab initio} calculations to calculate the energy of different collinear magnetic states on this planar structure under geometry optimizations (see details in the supplementary material). The result indicates that the most stable configuration exhibits the N\'eel antiferromagnetic order on Fe atoms while S atoms are non-magnetic, as shown in Fig.~\ref{Fig:FeS2}a.


Thus, this planar FeS$_2$ structure corresponds to the pentagonal lattice system discussed above. In particular, it exhibits the symmetries $\{C_{2\perp}||\mathcal{M}_{x}\}$, $\{C_{2\perp}||\mathcal{M}_{y}\}$, $\{C_{2\perp}||\mathcal{M}_{xy}\}$, and $\{C_{2\perp}||\mathcal{M}_{\bar{x}y}\}$, which relate two magnetic sublattices. The electronic band structure obtained from \emph{ab initio} calculations is shown in Fig.~\ref{Fig:FeS2}b, displaying features consistent with the tight-binding model result in Fig.~\ref{Fig:C4}a, that inverse spin-splitting occurs along the $\Gamma\mbox{-}C$ and $\Gamma\mbox{-}D$ paths, while spin-degeneracy is observed along the $\Gamma\mbox{-}X$, $\Gamma\mbox{-}Y$, and $\Gamma\mbox{-}M$ paths. These behaviors characterize planar pentagonal FeS$_2$ as a $g$-wave altermagnet as expected. Additionally, we present the spin density isosurfaces for bands with different levels of spin-splitting in Fig.~\ref{Fig:FeS2}c and Fig.~\ref{Fig:FeS2}d, where a larger spin-splitting coincides with a more anisotropic spin density. 

Next, we study the effect of strains along the diagonal directions on planar FeS$_2$, as indicated by the arrows in Fig.~\ref{Fig:FeS2}a. We introduce the strain in \emph{ab initio} calculations by slightly reducing the angle $\gamma$ between lattice vectors \textbf{a} and \textbf{b} to less than 90\degree. Fig.~\ref{Fig:FeS2}e presents the electronic band structure for $\gamma=88.0\degree$, where inverse spin-splitting appears along the $\Gamma\mbox{-}X\mbox{-}C\mbox{-}\Gamma$ and $\Gamma\mbox{-}Y\mbox{-}D\mbox{-}\Gamma$ paths, while the bands remain degenerate along the $\Gamma\mbox{-}M$ path, displaying the same features as in the model result shown in Fig.~\ref{Fig:C2}a. This behavior indicates a transition from $g$-wave to $d$-wave altermagnetism, as predicted by the tight-binding model analysis. 

\begin{figure*}[htbp]
\centering
\includegraphics[width=0.75\linewidth]{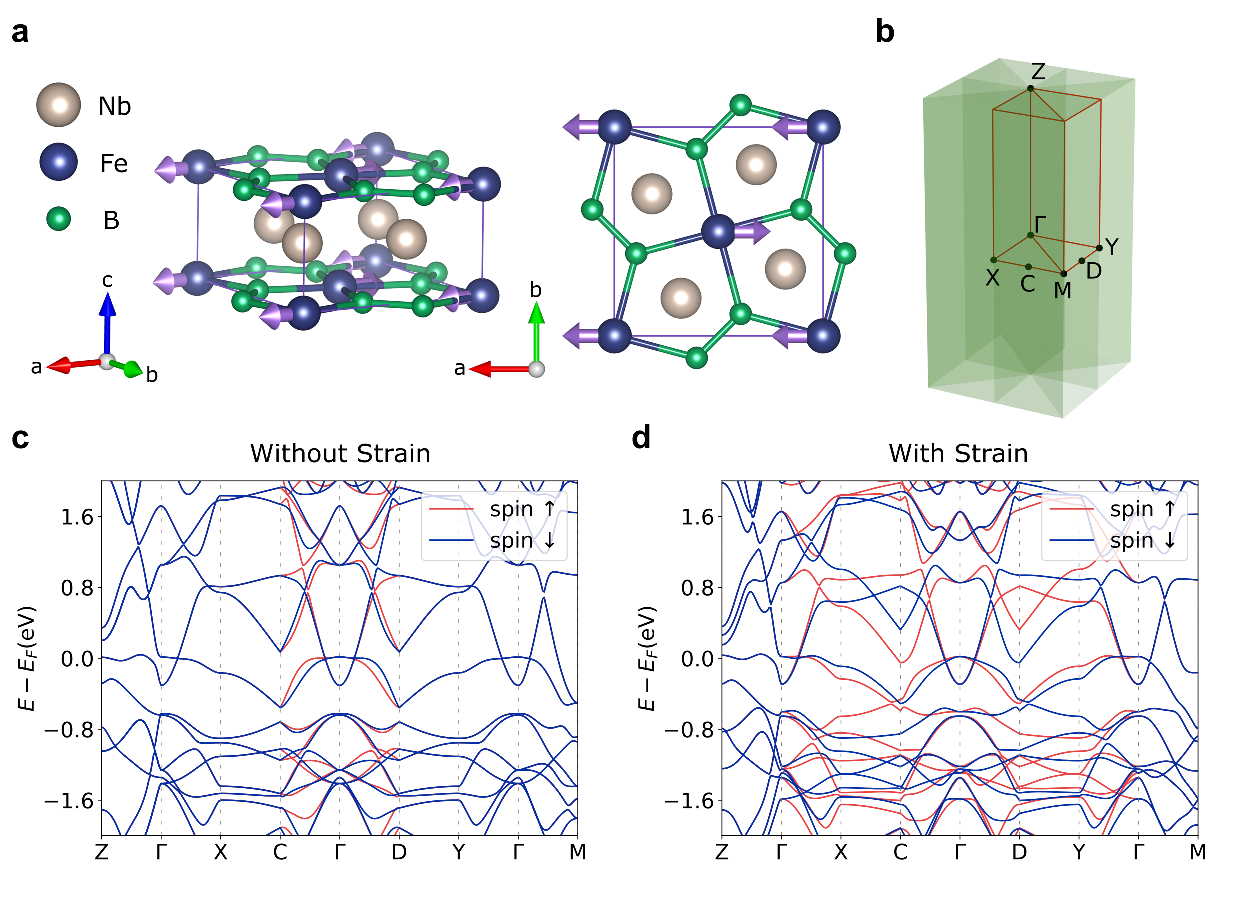}
\caption{\textbf{Crystal structure and electronic bands of Nb$_2$FeB$_2$.} \textbf{a} Crystal structure of Nb$_2$FeB$_2$ shown from two perspectives. The purple arrows indicate the directions of localized magnetic moments on Fe atoms. \textbf{b} The first Brillouin zone of Nb$_2$FeB$_2$ and high symmetry moment points of our interest. \textbf{c} and \textbf{d} The electronic band structures of Nb$_2$FeB$_2$ under two conditions: without diagonal strain (where the angle $\gamma$ between lattice parameters $a$ and $b$ is $90\degree$) and with diagonal strain (where 
$\gamma = 88\degree$).}
\label{Fig:NbFeB}
\end{figure*}

In addition to the two-dimensional cases, we highlight Nb$_2$FeB$_2$, a three-dimensional material predicted to exhibit $g$-wave altermagnetism~\cite{Nb2FeB21,Nb2FeB22}. Fig.~\ref{Fig:NbFeB}a shows the crystal structure of Nb$_2$FeB$_2$, which consists of alternate layers of Fe-B and Nb atoms along $c$ axis. Each Fe-B layer exhibits the same Cairo pentagonal structure discussed earlier, and the magnetic Fe atoms carry an in-plane N\'eel antiferromagnetic spin arrangement. Thus, the previous analysis of symmetry-induced altermagnetism also applies to Nb$_2$FeB$_2$. Its electronic band structures without and with diagonal strain are shown in Fig.~\ref{Fig:NbFeB}c and Fig.~\ref{Fig:NbFeB}d, where the angle $\gamma$ between lattice vectors \textbf{a} and \textbf{b} is 90\degree and 88\degree     respectively. These results demonstrate that Nb$_2$FeB$_2$ is a $g$-wave altermagnet and it undergoes a transition from $g$-wave to $d$-wave altermagnetism under diagonal strain, as expected. Furthermore, in both altermagnetic materials, the spin-splitting increases significantly after applying strain, consistent with the model analysis.\\

\noindent
\textbf{Conclusion and Discussion}
\\

Our study uncovers the first microscopic mechanism underlying $g$-wave altermagnetism in two dimension, realized in Cairo pentagonal lattice. Based on the symmetry analysis, we propose a simple but realistic tight-binding model containing both magnetic and non-magnetic sites that successfully realizes the altermagnetic band spin-splitting with $g$-wave symmetry. The lattice-spin symmetry ensured by the presence of non-magnetic atoms guarantees the existence of altermagnetism in this antiferromagnetically ordered pentagonal system. Our analysis of the dependence of spin-splitting on parameters in the tight-binding model provides one of the mechanisms, that both the hopping between magnetic site and non-magnetic site, and the hopping between non-magnetic sites are indispensable for the altermagnetic behavior. We note that there are other possible origins of altermagnetism, such as the anisotropic ordering of local orbitals, which are beyond our discussion in this work. 

Another important finding from our model is that applying strain to the crystal can induce a transformation in the type of altermagnetism. Strain reduces the original lattice-spin symmetry, introducing anisotropic hopping terms in the tight-binding model and making the spin-splitting behavior more complex. Near the $\Gamma$ point, we observe that strain applied along the diagonal direction of this pentagonal structure causes a transition from $g$-wave to $d$-wave altermagnetism. Additionally, we find that this pentagonal altermagnet contains spin-polarized nodal points protected by $\{\mathcal{T}C_{2\perp}||\mathcal{T}C_{2z}\}$ symmetry. Breaking this symmetry could lead to non-trivial topological bands and the quantum Hall effect for electrons.

Finally, we identify several candidate materials through \emph{ab initio} calculations, including monolayer FeS$_2$ and bulk Nb$_2$FeB$_2$, which show promise for exploring altermagnetism in Cairo pentagonal crystal structures. In particular, we examine the effect of diagonal strain on these crystals, observing that altermagnetism transforms from $g$-wave to $d$-wave as expected. These findings suggest new paradigm of tuning the type of altermagnetism by strain, opening new possibilities for enabling strain-tuned spintronic devices and the interplay with other degrees of freedom. 

\vspace{4mm}
\noindent
\MakeUppercase{\textbf{Methods}}\\
We use Vienna Ab initio Simulation Package (VASP)~\cite{PhysRevB.49.14251} for all density functional theory calculations. Projector-augmented-wave (PAW) potential was utilized for ion-electron interaction, and the generalized gradient approximation (GGA) with the Perdew-Burke-Ernzerhof (PBE) functional~\cite{PhysRevLett.77.3865} was applied to describe electron exchange-correlation interaction. The energy cutoff was set to be 600 eV. More details are in the supplementary material.

\vspace{4mm}
\noindent
\MakeUppercase{\textbf{Code Availability}}\\
{\small
The codes and scripts used for the numerical calculations reported in this paper are available from the first author (S.L.) upon request.
}\\

\noindent
\MakeUppercase{\textbf{Data Availability}}\\
{\small
The data analyzed in the present study is available from the first author (S.L.) upon request.
}

\bibliography{altermag}

\vspace{4mm}
\noindent
\uppercase{\textbf{Acknowledgements}}\\

This work is supported by the Center for Molecular Magnetic Quantum Materials, an Energy Frontier Research Center funded by the U.S. Department of Energy, Office of Science, Basic Energy Sciences under Award no. DE-SC0019330. Computations were done using the utilities of the University of Florida Research Computing.

\vspace{4mm}
\noindent
\uppercase{\textbf{Author contributions}}\\

C.J. and S.L. conceived, and C.J. supervised the work. S.L. performed calculations based on the tight-binding model. Y.Z., A.B., and X.Z. performed DFT calculations. S.L. analyzed the data. S.L., Y.Z., X.Z., and C.J. wrote the manuscript, with input from all authors. All authors reviewed the manuscript. 

\vspace{4mm}
\noindent
\uppercase{\textbf{Competing interests}}\\

The authors declare no competing interests.

\vspace{4mm}
\noindent
\uppercase{\textbf{Additional Information}}\\
{\small \textbf{Supplementary Information} The online version contains supplementary material available at \ldots\\

\noindent
\textbf{Correspondence} should be addressed to C.J. Request for materials should be directed to S.L..
}

\end{document}


\title{Supplementary Materials for ``Altermagnetism and Strain Induced Altermagnetic Transition in Cairo Pentagonal Monolayer"}

\author{Shuyi Li}
\affiliation{Department of Physics, University of Florida, Gainesville, FL 32611, USA}

\author{Yu Zhang}
\affiliation{Department of Physics, University of Florida, Gainesville, FL 32611, USA}

\author{Adrian Bahri}
\affiliation{Department of Physics, University of Florida, Gainesville, FL 32611, USA}

\author{Xiaoliang Zhang}
\affiliation{Department of Physics, University of Florida, Gainesville, FL 32611, USA}

\author{Chunjing Jia}
\affiliation{Department of Physics, University of Florida, Gainesville, FL 32611, USA}

\maketitle

\renewcommand{\theequation}{S\arabic{equation}}
\renewcommand{\thefigure}{S\arabic{figure}}
\renewcommand{\bibnumfmt}[1]{[S#1]}
\renewcommand{\citenumfont}[1]{S#1}

\section{Tight-Binding Model}
\begin{figure}[htbp]\label{supFig:model}
\centering
\includegraphics[width=0.5\linewidth]{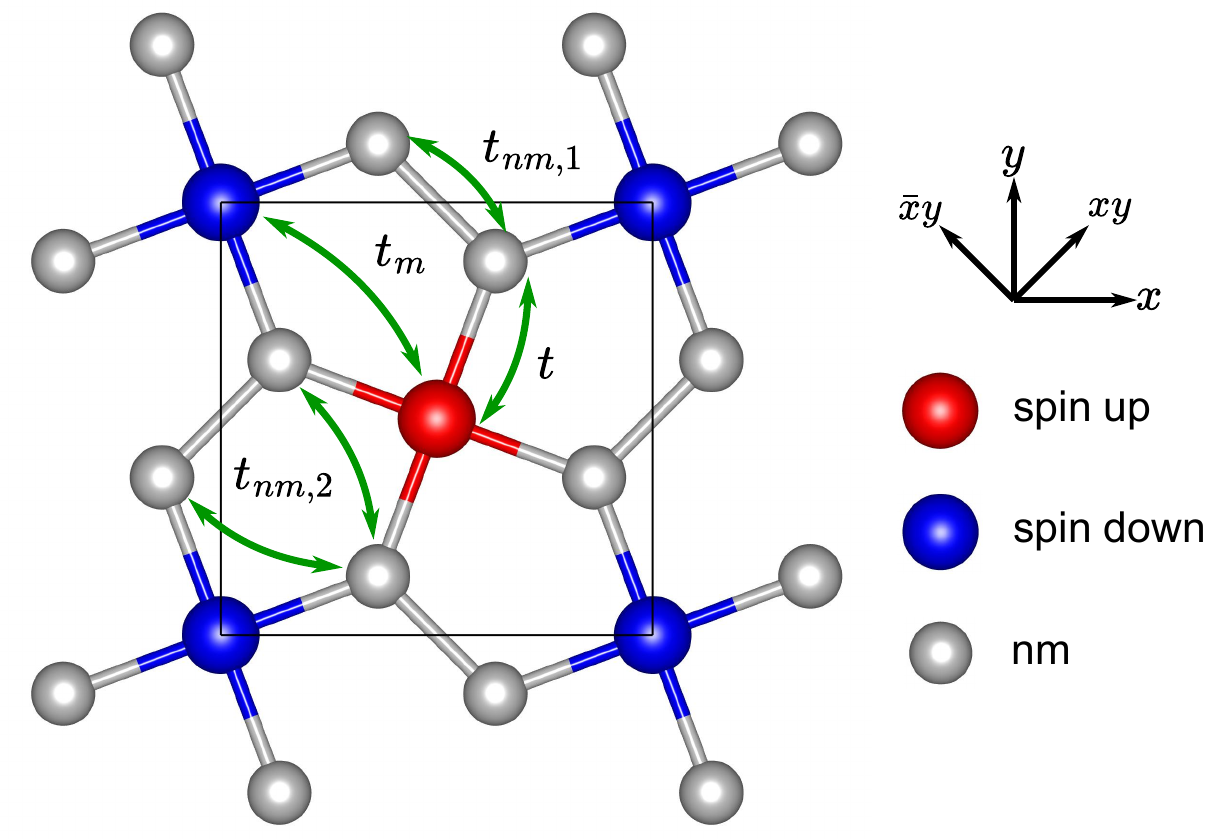}
\caption{A schematic of the pentagonal lattice with space group $P4/mbm$ and the tight-binding model on it.}
\label{Fig:appendix}
\end{figure}

The Cairo pentagonal lattice under consideration is shown in Fig.~\ref{supFig:model}. The lattice vectors are defined as $\mathbf{a}_1=(a,0)$ and $\mathbf{a}_2=(0,a)$, where $a$ is the lattice constant and set to be 1 for convenience. Each primitive unit cell contains two magnetic sites and four non-magnetic sites, their coordinates are
\begin{equation}
\begin{aligned}
    \text{magnetic}:&~(0,0),~(\frac{a}{2},\frac{a}{2})\\
    \text{non-magnetic}:&~(d,\frac{a}{2}-d),~(d-\frac{a}{2},d),~(-d,d-\frac{a}{2}),~(\frac{a}{2}-d,-d),
\end{aligned}
\end{equation}
where $0<d<a/2$. Without loss of generality, we set spin-up site at $(0,0)$, and $0<d=0.1a<a/4$ in our calculation.\\

The tight-binding Hamiltonian is
\begin{equation}\label{Sup-Eq-model}
\begin{aligned}
    H&=-(1+\delta)\sum_{\langle i,j\rangle_{xy}}t_{ij}c_{i\sigma}^{\dagger}c_{j\sigma}-(1-\delta)\sum_{\langle i,j\rangle_{\bar{x}y}}t_{ij}c_{i\sigma}^{\dagger}c_{j\sigma}-J\sum_{i\in\text{m},\sigma,\sigma'}\mathbf{S}_i\cdot c_{i\sigma}^{\dagger}\bm{\sigma}_{\sigma\sigma'}c_{i\sigma'}\\
    &+(\epsilon_{m}-\mu)\sum_{i\in m,\sigma}c_{i\sigma}^{\dagger}c_{i\sigma}+(\epsilon_{nm}-\mu)\sum_{i\in nm,\sigma}c_{i\sigma}^{\dagger}c_{i\sigma},
\end{aligned}
\end{equation}
where $\delta$ represents the strength of anisotropy caused by diagonal strains. We perform Fourier transformation into each electron operator
\begin{equation}\label{Fourier}
    c_{i\sigma}=\frac{1}{\sqrt{N}}\sum_{\mathbf{k}}c_{\mathbf{k}\sigma}e^{i\mathbf{r}_i\mathbf{k}}
\end{equation},
where $N$ is the number of unit cells. Then the Hamiltonian in Eq. (\ref{Sup-Eq-model}) becomes
\begin{equation}\label{Hk}
    H=\sum_{\mathbf{k},\sigma}\psi_{\mathbf{k}\sigma}^{\dagger}H_{\mathbf{k}\sigma}\psi_{\mathbf{k}\sigma},
\end{equation}
where $\psi_{\mathbf{k}\sigma}^{\dagger}=[c_{1\sigma\mathbf{k}}^{\dagger},\cdots, c_{6\sigma\mathbf{k}}^{\dagger}]$. The matrices $H_{\mathbf{k}\sigma}$ are written as
\begin{equation}
    H_{\mathbf{k}\uparrow}=
    \begin{bmatrix}
        \epsilon_m-JS & t_{AB} & t_{A\alpha} & t_{A\beta} & t_{A\gamma} & t_{A\eta}\\
        t_{AB}^* & \epsilon_m+JS & t_{B\alpha} & t_{B\beta} & t_{B\gamma} & t_{B\eta}\\
        t_{A\alpha}^* & t_{B\alpha}^* & \epsilon_{nm} & t_{\alpha\beta} & t_{\alpha\gamma} & t_{\alpha\eta}\\
        t_{A\beta}^* & t_{B\beta}^* & t_{\alpha\beta}^* & \epsilon_{nm} & t_{\beta\gamma} & t_{\beta\eta}\\
        t_{A\gamma}^* & t_{B\gamma}^* & t_{\alpha\gamma}^* & t_{\beta\gamma}^* & \epsilon_{nm} & t_{\gamma\eta}\\
        t_{A\eta}^* & t_{B\eta}^* & t_{\alpha\eta}^* & t_{\beta\eta}^* & t_{\gamma\eta}^* & \epsilon_{nm}\\
        \end{bmatrix}-\mu*\mathbbm{1}_{6\times6},
\end{equation}
and
\begin{equation}
    H_{\mathbf{k}\downarrow}=
    \begin{bmatrix}
        \epsilon_m+JS & t_{AB} & t_{A\alpha} & t_{A\beta} & t_{A\gamma} & t_{A\eta}\\
        t_{AB}^* & \epsilon_m-JS & t_{B\alpha} & t_{B\beta} & t_{B\gamma} & t_{B\eta}\\
        t_{A\alpha}^* & t_{B\alpha}^* & \epsilon_{nm} & t_{\alpha\beta} & t_{\alpha\gamma} & t_{\alpha\eta}\\
        t_{A\beta}^* & t_{B\beta}^* & t_{\alpha\beta}^* & \epsilon_{nm} & t_{\beta\gamma} & t_{\beta\eta}\\
        t_{A\gamma}^* & t_{B\gamma}^* & t_{\alpha\gamma}^* & t_{\beta\gamma}^* & \epsilon_{nm} & t_{\gamma\eta}\\
        t_{A\eta}^* & t_{B\eta}^* & t_{\alpha\eta}^* & t_{\beta\eta}^* & t_{\gamma\eta}^* & \epsilon_{nm}\\
        \end{bmatrix}-\mu*\mathbbm{1}_{6\times6},
\end{equation}
where the matrix elements are
\begin{equation}
    \begin{aligned}
        t_{AB}&=-2t_m(1+\delta)\cos(\frac{a}{2}k_x+\frac{a}{2}k_y)-2t_m(1-\delta)\cos(\frac{a}{2}k_x-\frac{a}{2}k_y),\\
        t_{A\alpha}&=-t(1-\delta)e^{i((\frac{a}{2}-d)k_x-dk_y)},\\
        t_{A\beta}&=-t(1-\delta)e^{i(-(\frac{a}{2}-d)k_x+dk_y)},\\
        t_{A\gamma}&=-t(1+\delta)e^{i(dk_x+(\frac{a}{2}-d)k_y)},\\
        t_{A\eta}&=-t(1+\delta)e^{i(-dk_x-(\frac{a}{2}-d)k_y)},\\
        t_{B\alpha}&=-t(1-\delta)e^{i(-dk_x+(-d+\frac{a}{2})k_y)},\\
        t_{B\beta}&=-t(1-\delta)e^{i(dk_x+(d-\frac{a}{2})k_y)},\\
        t_{B\gamma}&=-t(1+\delta)e^{i((d-\frac{a}{2})k_x-dk_y)},\\
        t_{B\eta}&=-t(1+\delta)e^{i((-d+\frac{a}{2})k_x+dk_y)},\\
        t_{\alpha\beta}&=-t_{nm,1}(1+\delta)e^{i(2dk_x+2dk_y)},\\
        t_{\alpha\gamma}&=-t_{nm,2}(1+\delta)e^{i((2d-\frac{a}{2})k_x-\frac{a}{2}k_y)}-t_{nm,2}(1-\delta)e^{i((2d-\frac{a}{2})k_x+\frac{a}{2}k_y)},\\
        t_{\alpha\eta}&=-t_{nm,2}(1+\delta)e^{i(-\frac{a}{2}k_x+(2d-\frac{a}{2})k_y)}-t_{nm,2}(1-\delta)e^{i(\frac{a}{2}k_x+(2d-\frac{a}{2})k_y)},\\
        t_{\beta\gamma}&=-t_{nm,2}(1+\delta)e^{i(\frac{a}{2}k_x+(\frac{a}{2}-2d)k_y)}-t_{nm,2}(1-\delta)e^{i(-\frac{a}{2}k_x+(\frac{a}{2}-2d)k_y)},\\
        t_{\beta\eta}&=-t_{nm,2}(1+\delta)e^{i((\frac{a}{2}-2d)k_x+\frac{a}{2}k_y)}-t_{nm,2}(1-\delta)e^{i((\frac{a}{2}-2d)k_x-\frac{a}{2}k_y)},\\
        t_{\gamma\eta}&=-t_{nm,1}(1-\delta)e^{i(-2dk_x+2dk_y)},\\
    \end{aligned}
\end{equation}
and $\mathbbm{1}_{6\times6}$ is a $6\times6$ identity matrix.

\section{Taylor Expansion of $E_{n,\sigma}(\mathbf{k})$ near $\Gamma$ point}
We investigate the behavior of band dispersions near $\Gamma$ point by performing Taylor expansion. Near $\mathbf{k}=\mathbf{0}$, we assume that an isolated band can be represented by
\begin{equation}
\begin{aligned}
    E_{\sigma}(\mathbf{k})&=a_{0,\sigma}+a_{21,\sigma}k_x^2+a_{22,\sigma}k_xk_y+a_{23,\sigma}k_y^2\\
    &+a_{41,\sigma}k_x^4+a_{42,\sigma}k_x^3k_y+a_{43,\sigma}k_x^2k_y^2+a_{44,\sigma}k_xk_y^3+a_{45,\sigma}k_y^4,
\end{aligned}
\end{equation}
there is no linear and cubic term because of inversion symmetry. Then we expand the characteristic equation
\begin{equation}
    \text{det}|E_{\sigma}\mathbbm{1}_{6\times6}-H_{\mathbf{k}\sigma}|=0
\end{equation}
at $\mathbf{k}=\mathbf{0}$ to solve the coefficients.\\

The model parameters used are $t_m=-0.4$, $t_{nm,1}=0.9$, $t_{nm,2}=0.6$, $J=1$, $\epsilon_m=\epsilon_{nm}=0$ and $\mu=0$. In the case of $\delta = 0$, there are four isolated bands near the $\Gamma$ point for each spin, the coefficients are in Table.~\ref{table:1}. The leading term of spin-splitting for the $n$th pair bands near the $\Gamma$ point is
\begin{equation}
    \Delta E_n(\mathbf{k})=2a_{n,42,\uparrow}k_xk_y(k_x^2-k_y^2),
\end{equation}
which is belong to $g$-wave symmetry.\\
\vspace{3mm}
\begin{table}
\begin{tabular}{ |c|c|c|c|c| } 
 \hline
 Band Index $n$ & 1 & 2 & 5 & 6 \\ 
 $a_{0,\uparrow}$ & -5.19 & 0.106 & 1.5 & 1.79 \\
 $a_{21,\uparrow}$ & 0.267 & -0.414 & 0.3 & 0.27 \\
 $a_{22,\uparrow}$ & 0 & 0 & 0 & 0 \\
 $a_{23,\uparrow}$ & 0.267 & -0.414 & 0.3 & 0.27 \\
 $a_{41,\uparrow}$ & -0.00425 & -0.0341 & -0.664 & 0.0497 \\
 $a_{42,\uparrow}$ & 0.000141 & -0.166 & 1.35 & -0.491 \\
 $a_{43,\uparrow}$ & -0.0111 & 0.631 & -0.551 & -0.00221 \\
 $a_{44,\uparrow}$ & -0.000141 & 0.166 & -1.35 & 0.491 \\
 $a_{45,\uparrow}$ & -0.00425 & -0.0341 & -0.664 & 0.0497 \\
 $a_{0,\downarrow}$ & -5.19 & 0.106 & 1.5 & 1.79 \\
 $a_{21,\downarrow}$ & 0.267 & -0.414 & 0.3 & 0.27 \\
 $a_{22,\downarrow}$ & 0 & 0 & 0 & 0 \\
 $a_{23,\downarrow}$ & 0.267 & -0.414 & 0.3 & 0.27 \\
 $a_{41,\downarrow}$ & -0.00425 & -0.0341 & -0.664 & 0.0497 \\
 $a_{42,\downarrow}$ & -0.000141 & 0.166 & -1.35 & 0.491 \\
 $a_{43,\downarrow}$ & -0.0111 & 0.631 & -0.551 & -0.00221 \\
 $a_{44,\downarrow}$ & 0.000141 & -0.166 & 1.35 & -0.491 \\
 $a_{45,\downarrow}$ & -0.00425 & -0.0341 & -0.664 & 0.0497 \\
 \hline
\end{tabular}
\caption{Coefficients of $E_{n,\sigma}(\mathbf{k})$ in the case of $\delta = 0$.}
\label{table:1}
\end{table}

In the case of $\delta = 0.1$, there are six isolated bands near the $\Gamma$ point for each spin, the coefficients are in Table. \ref{table:2}. The leading term of spin-splitting for the $n$th pair bands near the $\Gamma$ point is
\begin{equation}
    \Delta E_n(\mathbf{k})=(a_{n,21,\uparrow}-a_{n,23,\uparrow})(k_x^2-k_y^2),
\end{equation}
which is belong to $d$-wave symmetry.\\
\vspace{3mm}
\begin{table}
\begin{tabular}{ |c|c|c|c|c|c|c| } 
 \hline
 Band Index $n$ & 1 & 2 & 3 & 4 & 5 & 6 \\ 
 $a_{0,\uparrow}$ & -0.519 & 0.0821 & 0.81 & 0.99 & 1.4 & 1.91 \\
 $a_{21,\uparrow}$ & 0.266 & -0.416 & -0.228 & -0.0392 & 0.0625 & 0.355 \\
 $a_{22,\uparrow}$ & 0.0401 & -0.203 & 0.432 & -0.804 & 0.756 & -0.221 \\
 $a_{23,\uparrow}$ & 0.268 & -0.392 & -0.0054 & -0.894 & 1.01 & 0.0129 \\
 $a_{0,\downarrow}$ & -0.519 & 0.0821 & 0.81 & 0.99 & 1.4 & 1.91 \\
 $a_{21,\downarrow}$ & 0.268 & -0.392 & -0.0054 & -0.894 & 1.01 & 0.0129 \\
 $a_{22,\downarrow}$ & 0.0401 & -0.203 & 0.432 & -0.804 & 0.756 & -0.221 \\
 $a_{23,\downarrow}$ & 0.266 & -0.416 & -0.228 & -0.0392 & 0.0625 & 0.355 \\
 \hline
\end{tabular}
\caption{Coefficients of $E_{n,\sigma}(\mathbf{k})$ in the case of $\delta = 0.1$.}
\label{table:2}
\end{table}

\section{Band Degeneracy under $t_{nm,1}=0$ and $t_{nm,2}=0$}
The case of $t_{nm,1}=0$ is equivalent to that of a lattice with $d=a/4$, which is shown in Fig.~\ref{Fig:appendix}. This system clearly exhibits mirror symmetries $\mathcal{M}_x$ and $\mathcal{M}_y$. 

\begin{figure}[htbp]
\centering
\includegraphics[width=0.3\linewidth]{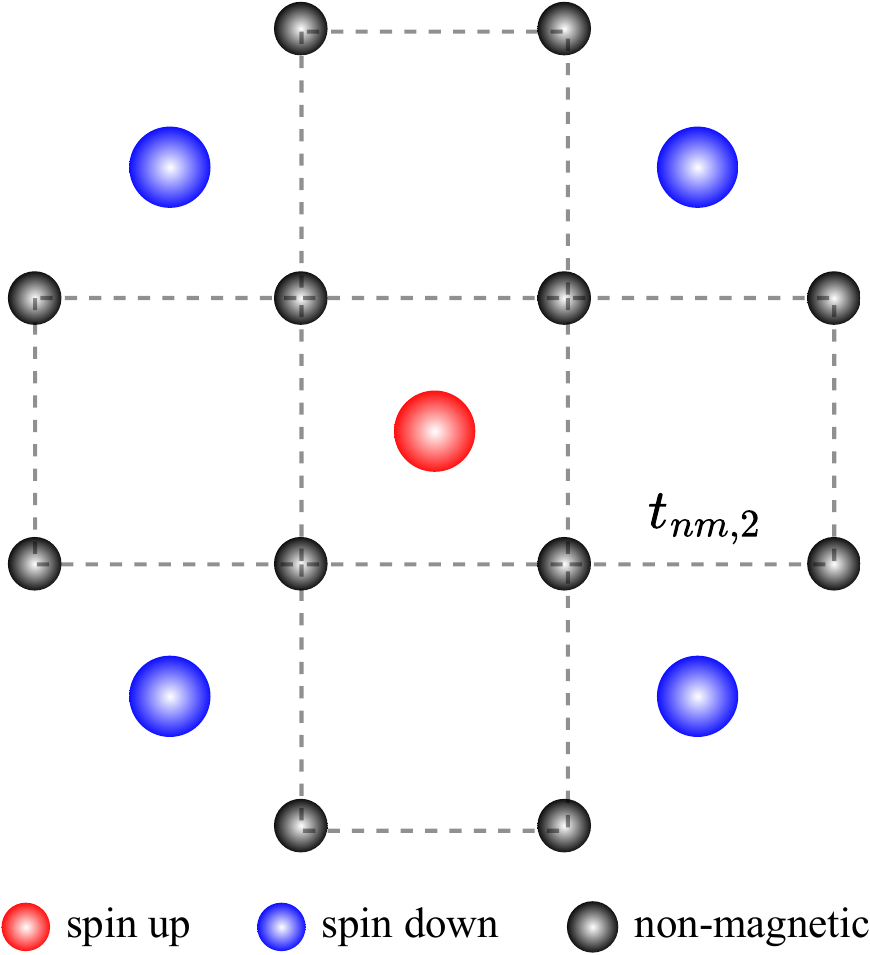}
\caption{The equivalent lattice of the Cairo pentagonal lattice when $t_{nm,1}=0$ in its tight-binding model.}
\label{Fig:appendix}
\end{figure}

The case of $t_{nm,2}=0$ is equivalent to that of a lattice with $d=0$. With this formula, it can be directly shown that the characteristic equations of $H_{kx,ky,\sigma}$, $H_{kx,-ky,\sigma}$, and $H_{-kx,ky,\sigma}$ are the same:  
\begin{equation}
\begin{aligned}
    \text{det}|E\mathbbm{1}_{6\times6}-H_{kx,ky,\sigma}|&=\text{det}|E\mathbbm{1}_{6\times6}-H_{kx,-ky,\sigma}|,\\
    \text{det}|E\mathbbm{1}_{6\times6}-H_{kx,ky,\sigma}|&=\text{det}|E\mathbbm{1}_{6\times6}-H_{-kx,ky,\sigma}|,\\
\end{aligned}
\end{equation}
which implies that $E_{n,\sigma}(kx,ky)=E_{n,\sigma}(kx,-ky)=E_{n,\sigma}(-kx,ky)$. The combination of these symmetries with $\{C_{2\perp}||\mathcal{M}_x\}$ and $\{C_{2\perp}||\mathcal{M}_y\}$ ensures that $E_{n,\uparrow}(\mathbf{k})=E_{n,\downarrow}(\mathbf{k})$.

\section{Berry Curvature}
After taking into account the symmetry breaking terms
\begin{equation}\label{app-Eq-H1}
    H_1=\sum_{i\in (m\uparrow),\sigma}\sum_{\bm{\xi}}(t_1c^{\dagger}_{i,\sigma}c_{i+\bm{\xi}\sigma}-t_1c^{\dagger}_{i\sigma}c_{i-\bm{\xi},\sigma})+h.c.
\end{equation}
and
\begin{equation}\label{app-Eq-H2}
    H_2=-\sum_{i\in (m\uparrow),\sigma}(-1)^{\sigma}\sum_{\bm{\xi}}i(t_2c^{\dagger}_{i,\sigma}c_{i+\bm{\xi}\sigma}+t_2c^{\dagger}_{i\sigma}c_{i-\bm{\xi},\sigma})+h.c.,
\end{equation}
we perform the same process as in Eq.(\ref{Fourier}) and Eq.(\ref{Hk}) to obtain the Hamiltonian matrix $H_{\mathbf{k}\sigma}^{total}$. Then we use Fukui's method to calculate Berry curvature $\Omega_{n,\sigma}(\mathbf{k})$ and Chern number $\mathcal{C}_{n,\sigma}$ numerically~\cite{fukui}, which is given by
\begin{equation}
\begin{aligned}
    &\Omega_{n,\sigma}(k_x+\frac{1}{2}\delta k_x,k_y+\frac{1}{2}\delta k_y)\\
    =&-\frac{1}{\delta k_x\delta k_y}\text{Arg}[\langle u_{n,\sigma}(k_x,k_y)|u_{n,\sigma}(k_x+\frac{1}{2}\delta k_x,k_y)\rangle\\
    &\langle u_{n,\sigma}(k_x+\frac{1}{2}\delta k_x,k_y)|u_{n,\sigma}(k_x+\frac{1}{2}\delta k_x,k_y+\frac{1}{2}\delta k_y)\rangle\\
    &\langle u_{n,\sigma}(k_x+\frac{1}{2}\delta k_x,k_y+\frac{1}{2}\delta k_y)|u_{n,\sigma}(k_x,k_y+\frac{1}{2}\delta k_y)\rangle\\
    &\langle u_{n,\sigma}(k_x,k_y+\frac{1}{2}\delta k_y)|u_{n,\sigma}(k_x,k_y)\rangle],\\
\end{aligned}
\end{equation}
\begin{equation}
    \mathcal{C}_{n,\sigma}=\frac{1}{2\pi}\sum_{\mathbf{k}}\Omega_{n,\sigma}(k_x+\frac{1}{2}\delta k_x,k_y+\frac{1}{2}\delta k_y)\delta k_x \delta k_y,
\end{equation}
where $|u_{n,\sigma}(k_x,k_y)\rangle$ is the $n$th eigenvector of the Hamiltonian $H_{\mathbf{k}\sigma}^{total}$. In the summation, we choose $\delta k_x=\delta k_y=2\pi/N_s$ with $N_s=201$.

\section{Details of \emph{Ab Initio} Calculations}
We use the Vienna Ab initio Simulation Package (VASP)~\cite{PhysRevB.49.14251} for all density-functional theory calculations. Projector-augmented-wave (PAW) potentials~\cite{PAW} were utilized for ion-electron interactions, and the generalized gradient approximation (GGA) with the Perdew-Burke-Ernzerhof (PBE) functional~\cite{PhysRevLett.77.3865} was applied to describe electron exchange-correlation interactions. The energy cutoff was set to 600 eV.

In the calculation of FeS$_2$, a $15\times15\times1$ Monkhorst-Pack $k$ point mesh was used and the GGA + U approach was adopted with $U_{eff}$ = 2.0 eV~\cite{Puthirath2021}. The monolayer FeS$_2$ is exfoliated from the bulk crystal in the (100) plane, forming a planar structure by manually relocating the S atoms. In our calculations, spin-orbit coupling is not included, resulting in the absence of magnetic anisotropy. 
To determine the magnetic ground state, we relaxed the structure under the ferromagnetic (FM) and Néel antiferromagnetic (AFM) configurations with 0.015 eV/$\si{\angstrom}$ and $10^{-6}$ eV as the convergence criterion for residual force and total-energy, respectively:
\begin{equation}
    E_{FM}=-29.138344~\text{eV},~E_{AFM}=-30.641527~\text{eV}.
\end{equation}
The results suggest that the N\'eel antiferromagnetic configuration has a lower energy with $\Delta$E = -1.50 eV. Keeping the lengths of lattice vectors unchanged, we apply shear strain in the diagonal direction by changing the angle $\gamma$ between lattice vectors $\mathbf{a}$ and $\mathbf{b}$ from 90\degree to 88\degree and relax the atomic positions.

For Nb$_2$FeB$_2$, a $8\times8\times13$ Monkhorst-Pack $k$-point mesh was used, and $U_{eff}=4.82$ eV \cite{Nb2FeB22} was set for strongly correlated Fe 3d electrons. We relax both lattice constant and atomic positions until the atomic force is less than 0.01eV/$\si{\angstrom}$.

\bibliography{altermagsup}